\documentclass{kapproc}

\normallatexbib

\begin{document}

\articletitle{The spectrum of boundary sine-Gordon theory
\thanks{Original title of the conference talk:
\textit{Sine-Gordon with Neumann boundary condition: the spectrum of
boundary states}
} 
}

\author{Z. Bajnok, L. Palla and \underbar{G. Tak{\'a}cs}
\thanks{Conference speaker}
}

\affil{Institute for Theoretical Physics, E{\"o}tv{\"o}s University,
Budapest, Hungary}

\begin{abstract}
We review our recent results on the on-shell description of
sine-Gordon model with integrable boundary conditions. We determined
the spectrum of boundary states by closing the boundary bootstrap and
gave a derivation of Al.B. Zamolodchikov's (unpublished) formulae for
the boundary energy and the relation between the Lagrangian
(ultraviolet) and bootstrap (infrared) parameters. These results have
been checked against numerical finite volume spectra coming from the
truncated conformal space approach. We find an entirely consistent
picture and strong evidence for the validity of the conjectured
spectrum and scattering amplitudes, which together give a complete
description of the boundary sine-Gordon theory on mass shell.
\end{abstract}


\section{\label{sec:introduction} Introduction}

In these proceedings we report on the results of our works
\cite{neumann,bajnok1,uvir}. Instead of following the line of the
original conference talk, we wish to give a summary of the results and
conjectures which together describe the on-shell spectral data of
boundary sine-Gordon theory.

Sine-Gordon field theory is one of the most important quantum field
theoretic models with numerous applications ranging from particle
theoretic problems to condensed matter systems, and one which has
played a central role in our understanding of \( 1+1 \) dimensional
field theories. A crucial property of the model is integrability,
which permits an exact analytic determination of many of its physical
properties and characteristic quantities. Integrability can also be
preserved in the presence of boundaries \cite{Skl}; for sine-Gordon
theory, the most general boundary potential that preserves
integrability was found by Ghoshal and Zamolodchikov
\cite{ghoshal}. They also introduced the notion of \lq boundary
crossing unitarity', and combining it with the boundary version of the
Yang-Baxter equations they were able to determine soliton reflection
factors on the ground state boundary; later Ghoshal completed this
work by determining the breather reflection factors \cite{ghoshal1}
using a boundary bootstrap equation first proposed by Fring and
K{\"o}berle \cite{FK}.

The first (partial) results on the spectrum of the excited
boundary states were obtained by Saleur and Skorik for Dirichlet
boundary conditions \cite{skorik}. However, they did not take into
account the boundary analogue of the Coleman-Thun mechanism, the
importance of which was first emphasized by Dorey et
al. \cite{bct}. Using this mechanism Mattsson and Dorey were able to
close the bootstrap in the Dirichlet case and determine the complete
spectrum and the reflection factors on the excited boundary states
\cite{mattsson}. Recently we used their ideas to obtain the spectrum
of excited boundary states and their reflection factors for the
Neumann boundary condition \cite{neumann} and then for the general
two-parameter family of integrable boundary conditions \cite{bajnok1}.

Another interesting problem is the relation between the ultraviolet
(UV) parameters that appear in the perturbed CFT Hamiltonian and the
infrared (IR) parameters in the reflection factors. This relation was
first obtained by Al. B. Zamolodchikov \cite{unpublished} together
with a formula for the boundary energy; however, his results remain
unpublished. In order to have these formulae, we rederived them in our
paper \cite{uvir}, where we used them to check the consistency of the
spectrum and of the reflection factors against a boundary version of
truncated conformal space approach (TCSA). Combining the TCSA results
with analytic methods of the Bethe Ansatz we found strong evidence
that our understanding of the spectrum of boundary sine-Gordon model
is indeed correct.

\section{\label{sec:boundary_bootstrap} Boundary bootstrap in sine-Gordon theory }

Boundary sine-Gordon theory is defined by the action
\begin{eqnarray}
\mathcal{A}_{sG}&=&\int ^{\infty }_{-\infty }dt\Bigg( \int
^{0}_{-\infty }dx\left[ \frac{1}{2}\partial _{\mu }\Phi \partial ^{\mu
}\Phi +\frac{m^{2}_{0}}{\beta ^{2}}\cos \beta \Phi \right] \nonumber\\ 
&+&M_{0}\cos \frac{\beta }{2}\left( \Phi (0,\, t)-\phi _{0}\right) \Bigg) 
\label{bsg_action}
\end{eqnarray}
where \( \Phi (x,\, t) \) is a real scalar field and \( M_0\),
\(\phi_0\) are the two parameters characterizing the boundary
condition:
\begin{equation}
\label{hatfel}
\partial_x\Phi(x,t)\vert_{x=0}=-M_0\frac{\beta}{2}\sin\left(\frac{\beta}{2}
(\Phi(0,t)-\phi_0)\right).
\end{equation}
Ghoshal and Zamolodchikov showed that the above model is integrable
\cite{ghoshal} and that the boundary term is the most general
consistent with integrability and containing no time derivatives of
the field $\Phi$.

In the bulk sine-Gordon model the particle spectrum consists of the
soliton \( s \), the antisoliton \( \bar{s} \), and the breathers \(
B^{n} \), which appear as bound states of a soliton and an
antisoliton. As a consequence of the integrable nature of the model any
scattering amplitude factorizes into a product of two particle
scattering amplitudes which were found by Zamolodchikov and
Zamolodchikov \cite{ZZ}. Factorization of the scattering carries over
to the situation with integrable boundary conditions as well
\cite{ghoshal}.

\subsection{Ground state reflection factors}

In the presence of boundary, the bulk $S$-matrix must be supplemented
with the reflection amplitudes describing the interaction of the
particles with the boundary in order to specify the scattering theory
completely.  The most general reflection factor - modulo CDD-type
factors - of the soliton antisoliton multiplet \( |s,\bar{s}\rangle \)
on the ground state boundary, denoted by \( |\, \rangle \), satisfying
the boundary versions of the Yang-Baxter, unitarity and crossing
equations was found by Ghoshal and Zamolodchikov \cite{ghoshal}:
\begin{eqnarray}
R(\eta ,\vartheta ,u) & = & \left( \begin{array}{cc}
P^{+}(\eta ,\vartheta ,u) & Q(\eta ,\vartheta ,u)\\
Q(\eta ,\vartheta ,u) & P^{-}(\eta ,\vartheta ,u)
\end{array}\right) \nonumber \\
 & = & \left( \begin{array}{cc}
P_{0}^{+}(\eta ,\vartheta ,u) & Q_{0}(u)\\
Q_{0}(u) & P_{0}^{-}(\eta ,\vartheta ,u)
\end{array}\right) 
R_{0}(u)\frac{\sigma (\eta ,u)}{\cos (\eta )}\frac{\sigma (i\vartheta ,u)}{\cosh (\vartheta )}\, \, \, ,\nonumber \\
P_{0}^{\pm }(\eta ,\vartheta ,u) & = & \cos (\lambda u)\cos (\eta )\cosh (\vartheta )\mp \sin (\lambda u)\sin (\eta )\sinh (\vartheta )\nonumber \\
Q_{0}(u) & = & -\sin (\lambda u)\cos (\lambda u)\label{Rsas} 
\end{eqnarray}
where we introduced 
\begin{equation}
\label{lalam} \lambda =\frac{8\pi }{\beta ^{2}}-1\, ,
\end{equation} 
\(u=-i\theta\) denote the purely imaginary rapidity as in
\cite{ghoshal}, \( \eta \) and \( \vartheta \) are two real parameters
characterizing the solution,
\[ 
R_{0}(u)=\prod ^{\infty }_{l=1}\left[ \frac{\Gamma
(4l\lambda -\frac{2\lambda u}{\pi })\Gamma (4\lambda
(l-1)+1-\frac{2\lambda u}{\pi })}{\Gamma ((4l-3)\lambda
-\frac{2\lambda u}{\pi })\Gamma ((4l-1)\lambda +1-\frac{2\lambda
u}{\pi })}/(u\to -u)\right] 
\] 
is the boundary condition independent
part and 
\begin{eqnarray} 
&&\sigma (x,u)=\frac{\cos x}{\cos (x+\lambda u)}
\nonumber\\
&&\prod
^{\infty }_{l=1}\left[ \frac{\Gamma (\frac{1}{2}+\frac{x}{\pi
}+(2l-1)\lambda -\frac{\lambda u}{\pi })\Gamma
(\frac{1}{2}-\frac{x}{\pi }+(2l-1)\lambda -\frac{\lambda u}{\pi
})}{\Gamma (\frac{1}{2}-\frac{x}{\pi }+(2l-2)\lambda -\frac{\lambda
u}{\pi })\Gamma (\frac{1}{2}+\frac{x}{\pi }+2l\lambda -\frac{\lambda
u}{\pi })}/(u\to -u)\right] 
\nonumber
\end{eqnarray}
describes the boundary condition dependence. Minimality (i.e. minimal pole
structure) restricts $0\leq\eta\leq\pi(\lambda+1)/2$, while the
independent values of $\vartheta$ are $0\leq\vartheta\leq\infty$. As
it can be seen from the UV-IR relation to be discussed later
(eqn. (\ref{UV_IR_relation})), this covers exactly the range of
parameters in the Lagrangian description; therefore it is thought that
only the minimal solution is realized in boundary sine-Gordon
model. This is also confirmed by our TCSA analysis (see Section 4).

As a consequence of the bootstrap equations \cite{ghoshal} the breather reflection
factors share the structure of the solitonic ones, \cite{ghoshal1}: 
\begin{equation}
\label{Rbr1}
R^{(n)}(\eta ,\vartheta ,u)=R_{0}^{(n)}(u)S^{(n)}(\eta ,u)S^{(n)}(i\vartheta ,u)\, \, \, ,
\end{equation}
where 
\begin{eqnarray}
R_{0}^{(n)}(u)&=&\frac{\left( \frac{1}{2}\right) \left(
\frac{n}{2\lambda }+1\right) }{\left( \frac{n}{2\lambda
}+\frac{3}{2}\right) }\prod ^{n-1}_{l=1}\frac{\left( \frac{l}{2\lambda
}\right) \left( \frac{l}{2\lambda }+1\right) }{\left(
\frac{l}{2\lambda }+\frac{3}{2}\right) ^{2}}\nonumber \\ 
S^{(n)}(x,u)&=&\prod ^{n-1}_{l=0}\frac{\left( \frac{x}{\lambda \pi
}-\frac{1}{2}+\frac{n-2l-1}{2\lambda }\right) }{\left(
\frac{x}{\lambda \pi }+\frac{1}{2}+\frac{n-2l-1}{2\lambda }\right) }\,
,\quad
(x)=\frac{\sin \left( \frac{u}{2}+\frac{x\pi }{2}\right) }{\sin \left(
\frac{u}{2}-\frac{x\pi }{2}\right) }\, .
\label{Rbr2}
\end{eqnarray}
In general \( R_{0}^{(n)} \) describes the boundary independent properties
and the other factors give the boundary dependent ones.

\subsection{The spectrum of boundary bound states and the associated reflection factors}

In the general case, the spectrum of boundary bound states was derived in \cite{bajnok1}.
It is a straightforward generalization of the spectrum in the Dirichlet limit
previously obtained by Mattsson and Dorey \cite{mattsson}. The states can be
labeled by a sequence of integers \( |n_{1},n_{2},\dots ,n_{k}\rangle \).
Such a state exists whenever the \[
\frac{\pi }{2}\geq \nu _{n_{1}}>w_{n_{2}}>\nu _{n_{3}}>w_{n_{4}}>\dots \geq 0\]
 condition holds, where 
\[
\nu _{n}=\frac{\eta }{\lambda }-\frac{(2n+1)\pi }{2\lambda }\; 
\quad\mathrm{and}\;\quad  w_{n}=\pi -\frac{\eta }{\lambda }-\frac{(2n-1)\pi }{2\lambda }\; ,\]
denote the location of certain poles in $\sigma (\eta ,u)$. 
The mass of such a state (i.e. its energy above the ground state) is 

\begin{equation}
m_{|n_{1},n_{2},\dots ,n_{k}\rangle }=M\sum _{i\textrm{
}\mathrm{odd}}\cos (\nu _{n_{i}})+M\sum _{i\textrm{
}\mathrm{even}}\cos (w_{n_{i}})\, \, \, .\label{excited_energy}
\end{equation}
The reflection factors of the various particles on these boundary
states 
depend on whether $k$ is even or odd. When \( k \) is even, we have\[
Q_{|n_{1},n_{2},\dots ,n_{k}\rangle }(\eta ,\vartheta ,u)=Q(\eta ,\vartheta ,u)\prod _{i\textrm{ }\mathrm{odd}}a_{n_{i}}(\eta ,u)\prod _{i\textrm{ }\mathrm{even}}a_{n_{i}}(\bar{\eta },u)\, \, \, ,\]
and \[
P^{\pm }_{|n_{1},n_{2},\dots ,n_{k}\rangle }(\eta ,\vartheta ,u)=P^{\pm }(\eta ,\vartheta ,u)\prod _{i\textrm{ }\mathrm{odd}}a_{n_{i}}(\eta ,u)\prod _{i\textrm{ }\mathrm{even}}a_{n_{i}}(\bar{\eta },u)\, \, \, ,\]
for the solitonic processes, where\[
a_{n}(\eta ,u)=\prod _{l=1}^{n}\left\{ 2\left( \frac{\eta }{\pi }-l\right) \right\} \quad ;\quad \bar{\eta }=\pi (\lambda +1)-\eta \, \, \, .\]
and 
\[
\{y\}=\frac{\left( \frac{y+1}{2\lambda }\right) \left( \frac{y-1}{2\lambda }\right) }{\left( \frac{y+1}{2\lambda }-1\right) \left( \frac{y-1}{2\lambda }+1\right) }\]

For the breather reflection factors the analogous formula is 
\begin{equation}
\label{Rbe1}
R^{(n)}_{|n_{1},n_{2},\dots ,n_{k}\rangle }(\eta ,\vartheta
,u)=R^{(n)}(\eta ,\vartheta ,u)\prod _{i\textrm{
}\mathrm{odd}}b^{n}_{n_{i}}(\eta ,u)\prod _{i\textrm{
}\mathrm{even}}b^{n}_{n_{i}}(\bar{\eta },u)\, \, \, 
\end{equation}
where now 
\begin{equation}
b_{k}^{n}(\eta ,u)=\prod _{l=1}^{\min (n,k)}\left\{ \frac{2\eta }{\pi
}-\lambda +n-2l\right\}
\left\{ \frac{2\eta }{\pi }+\lambda
-n-2(k+1-l)\right\} \, \, \, .
\label{Rbe2}
\end{equation}
In the case when \( k \) is odd, the same formulae apply if in
the \(P^\pm\), \(Q\) and \(R^{(n)}\) 
ground state reflection factors the \( \eta \leftrightarrow \bar{\eta } \)
and \( s\leftrightarrow \bar{s} \) changes are made. 

\subsection{Closure of the bootstrap}

In our papers \cite{neumann,bajnok1} we presented an argument that the
bootstrap closes for the above spectrum. The essential steps are:

\begin{enumerate}
\item{}We conjectured the minimal spectrum (i.e. the states that are
necessary to include) by examining the reflection amplitudes of the
solitons.
\item{}We proved that these states must be included in the spectrum,
i.e. that the poles in the reflection factors corresponding to them
cannot be explained by any boundary Coleman-Thun diagram.
\item{}For all other poles of the breather and soliton reflection
factors we found an explanation in terms of one of the states listed
above or at least one boundary Coleman-Thun type diagram which had the
same order as the pole.
\end{enumerate}
The only thing that remains is to check that the full residues of
the poles can indeed be obtained as sums of contributions of all
possible diagrams, using only the states in the minimal spectrum. We
checked this in some of the simplest cases explicitly
\cite{neumann}. However, finding all the diagrams and computing all
the residues is a horrendous task, which we have not completed. From
TCSA we have overwhelming evidence that the spectrum and the
reflection factors are correct and we briefly discuss this evidence in
the sequel.

For the case of the Neumann boundary conditions \cite{neumann} we
noted that the conjectured spectrum implies that there are poles (in
breather reflection factors) whose residue can only be explained by
including contributions both from a boundary excited state and from a
Coleman-Thun type diagram. In boundary Lee-Yang model, a very similar
phenomenon was discussed by Dorey, Tateo and Watts \cite{bct}. There
it was also related to the fact that the closure of the bootstrap was
not unique. In the case of the sine-Gordon theory, however, the
phenomenon that a pole can only be explained by a combination of some
Coleman-Thun diagram together with some boundary excited state,
happens only for some special values of the parameter $\eta$ and so we
do not think that it is an indication of any nonuniqueness in the
bootstrap. Indeed, for generic values of the parameters the bootstrap
closure does seem to be uniquely determined and therefore we think
that even for the special values the correct closure of the bootstrap
is the one above, since we expect that the spectrum depends smoothly on
the parameters $\eta$ and $\vartheta$.

\section{Zamolodchikov's formulae}

Recently, Al. B. Zamolodchikov presented \cite{unpublished} a formula
for the relation between the UV and the IR parameters in the
boundary sine-Gordon model. We shall consider boundary sine-Gordon
theory as a joint bulk and boundary perturbation of the \( c=1 \) free
boson with Neumann boundary conditions (perturbed conformal field
theory, pCFT):
\begin{eqnarray}
\mathcal{A}_{pCFT}&=&\mathcal{A}^{N}_{c=1}+\mu \, \int ^{\infty
}_{-\infty }dt\int ^{0}_{-\infty }dx\, :\cos \beta \Phi (x,\, t):
\nonumber \\
&+&\tilde{\mu }\, \int ^{\infty }_{-\infty }dt\, :\cos \frac{\beta }{2}\left( \Phi (0,\, t)-\phi _{0}\right) :
\label{pCFT_action}
\end{eqnarray}
where the colons denote the standard CFT normal ordering, which defines the
normalization of the operators and of the coupling constants. The couplings
have nontrivial dimensions: \( [\mu ]=[ {\rm mass} ]^{2-\beta^2/4\pi}
\), \( [\tilde{\mu }]=[ {\rm mass}]^{1-\beta^2/8\pi}\).

The UV parameters associated to the boundary are $\tilde\mu$ and
$\phi_0$, while the IR parameters are $\eta$ and $\vartheta$ appearing
in (\ref{Rsas}). The other UV parameter $\mu$ is related to the
soliton mass $M$ the same way as for the bulk theory \cite{massgap}.
With the above conventions the UV-IR relation
\footnote{A similar relation was derived by Corrigan and Taormina
\cite{Ed} for sinh-Gordon theory, however, their normalization of the
coupling constants is different from the one natural in the perturbed
CFT framework.
} is
\begin{eqnarray}
\label{UV_IR}
\cos \left( \frac{\beta ^{2}\eta }{8\pi }\right) \cosh \left(
\frac{\beta ^{2}\vartheta }{8\pi }\right) & = &
\frac{\tilde\mu}{\sqrt{2\mu}}\sqrt{\sin\left(\frac{\beta ^{2}}{8}\right)}
\cos \left( \frac{\beta \phi _{0}}{2}\right)\,,
\nonumber \\ 
\sin \left( \frac{\beta ^{2}\eta }{8\pi }\right) \sinh
\left( \frac{\beta ^{2}\vartheta }{8\pi }\right) & = &
\frac{\tilde\mu}{\sqrt{2\mu}}\sqrt{\sin\left(\frac{\beta^{2}}{8}\right)}
\sin \left(\frac{\beta \phi _{0}}{2}\right)\,.
\label{UV_IR_relation} 
\end{eqnarray}
Zamolodchikov also gave the boundary energy as
\begin{eqnarray}
E(\eta ,\vartheta )=&-&\frac{M}{2\cos \frac{\pi }{2\lambda }}
\Big( 
\cos \left( \frac{\eta }{\lambda }\right) +\cosh \left(
\frac{\vartheta }{\lambda }\right)
-\frac{1}{2}\cos \left( \frac{\pi }{2\lambda }\right) 
\nonumber\\
&+&\frac{1}{2}\sin \left( \frac{\pi }{2\lambda }\right)
-\frac{1}{2}\Big)\, . 
\label{boundary_energy}
\end{eqnarray}

The above formula for the boundary energy can be derived using the
thermodynamic Bethe Ansatz for sinh-Gordon theory. The derivation,
however, contains some nontrivial analytic continuation due to the
fact that the sinh-Gordon TBA has no plateau solution in the
ultraviolet limit \cite{sinhG_tba}. For more details see \cite{uvir}, where we
also gave a derivation of the UV-IR relation from the exact vacuum
expectation values of boundary fields conjectured by Lukyanov,
Zamolodchikov and Zamolodchikov \cite{FZZ}.

We can perform a check on eqn. (\ref{boundary_energy}) which relates
the boundary energy obtained from TBA to the bootstrap spectrum,
thereby showing their consistency. It was noted in \cite{mattsson}
(for Dirichlet boundary condition) and in \cite{bajnok1} (for the
general case) that continuing analytically
\[
\eta\,\rightarrow\,\pi(\lambda+1)-\eta
\]
the roles of the boundary ground state \( |\rangle \) and of the boundary
first excited state \( |0 \rangle \) are interchanged. Therefore we
can calculate the energy difference between these two states from the
formula for the boundary energy, eqn. (\ref{boundary_energy}). The
result is 
\[
E(\pi(\lambda+1)-\eta,\vartheta)-E(\eta,\vartheta)=
M\cos\left(\frac{\eta}{\lambda}-\frac{\pi}{2\lambda}\right)
\]
which exactly equals the prediction of the bootstrap, i.e. 
\[
E_{ |0 \rangle }-E_{ |\rangle }=
M\cos\nu_0
\]
that follows from eqn. (\ref{excited_energy}). In our paper
\cite{uvir} we also checked that (\ref{boundary_energy}) was
consistent with other information known from previous literature, like
the boundary energy of the Lee-Yang model \cite{dptw} and the boundary
energy for sine-Gordon with Dirichlet boundary conditions
\cite{leclair}.

\section{TCSA verification}

Truncated Conformal Space is a method to compute the spectrum of a
perturbed CFT in finite volume. It works by computing the matrix
elements of the Hamiltonian in the basis of the conformal states and
then truncating the space of states to a finite dimensional subspace
by imposing a cut-off in the conformal weight. It was introduced by
Yurov and Zamolodchikov for the Lee-Yang model with periodic boundary
conditions \cite{YZ} and later extended to perturbations of $c=1$
theories \cite{frt} and to theories with boundary \cite{dptw}. As the
boundary sine-Gordon model is a perturbed $c=1$ CFT with boundary
conditions, we could use these developments to compute the spectrum on
a space-time strip, where space was an interval with (not necessarily
identical) integrable boundary conditions on both ends. For a description
of the technical details see \cite{neumann}.

On the other hand, one can make predictions for the finite size
spectrum starting from the knowledge of the bootstrap spectrum and
reflection factors and using a Bethe Ansatz technique.

The detailed results are described in
\cite{neumann} for the case of Neumann boundary condition (\(M_0=0\)
in eqn. (\ref{bsg_action})) and in \cite{uvir} for Dirichlet
(\(M_0\,\rightarrow\,\infty\)) and general boundary conditions. Here
we only describe the main conclusions.

\begin{enumerate}
\item{}The formulae (\ref{UV_IR_relation},\ref{boundary_energy}) and
the ground state reflection factors (\ref{Rsas},\ref{Rbr1}) are in
very good agreement with the numerical data.
\item{}Boundary excited states can be obtained as analytic
continuation of certain one-particle states in the framework of Bethe
Ansatz \cite{neumann}. The spectrum of the states that are accessible
in TCSA matches precisely with the one conjectured from the bootstrap.
\item{}If a pole in the reflection factors was explained by some
boundary Coleman-Thun diagram and has no corresponding bound state in
the bootstrap, there is no such state in the TCSA spectrum
either. This makes us confident that the rules used to draw these
diagrams are indeed correct, which is very important as for the time
being they have no field theoretical derivation.  
\end{enumerate}

\section{Conclusions}
We reported on work leading to a complete on-shell description of
boundary sine-Gordon theory using bootstrap methods. We also derived
Zamolodchikov's formulae for the boundary energy and the UV-IR
relation and compared the results to numerical TCSA calculations. We
found an excellent agreement and confirmed the general picture that
was formed of boundary sine-Gordon theory in the previous literature.

The main open problems are the calculation of off-shell quantities
(e.g. correlation functions) and exact finite size spectra. While
correlation functions in general present a very hard problem even in
theories without boundaries, in integrable theories significant
progress was made using form factors (for one-point functions of bulk
operators see e.g. \cite{bulk_vevs}). In addition, the vacuum
expectation values of boundary operators in sine-Gordon theory are
also known exactly \cite{FZZ}.  It would be interesting to make
further progress in this direction.

Concerning finite size spectra, there is already a version of the so-called
nonlinear integral equation for the vacuum (Casimir) energy with Dirichlet boundary
conditions \cite{leclair}, but it is not yet clear how to extend it to describe
excited states and more general boundary conditions as well, which also seems
to be a fascinating problem. 

It would also be interesting to work out a formalism (an analogue of
the Cutkosky rules of quantum field theory in the bulk) in which the
rules for the boundary Coleman-Thun diagrams can be justified.

\begin{acknowledgments}
We would like to thank P. Dorey, G. Watts and especially Al.B. Zamolodchikov
for very useful discussions. G.T. was supported by a Magyary postdoctoral fellowship
from the Hungarian Ministry of Education. This research was supported in part
by the Hungarian Ministry of Education under FKFP 0178/1999, 0043/2001 and by
the Hungarian National Science Fund (OTKA) T029802/99.
\end{acknowledgments}

\end{document}